\documentclass[a4paper,12pt]{article}
\usepackage{amsfonts,amsthm,amsmath,amssymb,graphicx,hyperref,youngtab}
\usepackage[bulletsep]{collref}

\def\Tr{{\rm Tr\, }}

\def\half{ {1 \over 2} }

\newcommand{\be}{\begin{equation}}
\newcommand{\bea}{\begin{eqnarray}}
\newcommand{\ee}{\end{equation}}
\newcommand{\eea}{\end{eqnarray}}

\newcommand{\AdS}{\mathrm{AdS}}
\newcommand{\Srm}{\mathrm{S}}
\newcommand{\diff}{\mathrm{d}}
\newcommand{\Rcal}{\mathcal{R}}

\begin{document}

\makeatletter
\@addtoreset{equation}{section}
\makeatother
\renewcommand{\theequation}{\thesection.\arabic{equation}}

\rightline{WITS-CTP-093}
\vspace{1.8truecm}

\vspace{15pt}

%%%%%%%%%%%%%%%%%

{\LARGE{   
\centerline{   \bf Extremal vs. Non-Extremal Correlators} 
\centerline {\bf  with Giant Gravitons} 
}}  

\vskip.9cm 

\thispagestyle{empty} \centerline{
    {\large \bf Pawel Caputa${}^{a,} $\footnote{ {\tt pawel.caputa@wits.ac.za}}, 
Robert de Mello Koch${}^{a,} $\footnote{ {\tt robert@neo.phys.wits.ac.za}}}
   {\large \bf and Konstantinos Zoubos${}^{b,}$\footnote{ {\tt kzoubos@nbi.dk}}   }
                                                       }

\vspace{.4cm}
\centerline{{\it ${}^a$ National Institute for Theoretical Physics}}
\centerline{{\it Department of Physics and Centre for Theoretical Physics }}
\centerline{{\it University of Witwatersrand, Wits, 2050, } }
\centerline{{\it South Africa } }

\vspace{.4cm}
\centerline{{\it ${}^b$ Niels Bohr Institute}}
\centerline{{ \it Blegdamsvej 17, DK-2100 Copenhagen},} 
\centerline{{ \it Denmark}} 
\vspace{1.4truecm}

%%%%%%%%%%%%%%%%%
\thispagestyle{empty}

\centerline{\bf Abstract}

\vskip.4cm 

We consider extremal and non-extremal three-point functions of two giant gravitons and one point-like 
graviton using Schur polynomials in ${\cal N}=4$ super Yang-Mills theory and holographically, using
a semiclassical Born-Infeld analysis as well as bubbling geometries. 
For non-extremal three-point functions our computations using all three approaches are in perfect agreement.
For extremal correlators we find that our results from the bubbling geometry analysis agree
with existing results from the gauge theory. The semiclassical Born-Infeld computation for
the extremal case is known to give a different answer, which we interpret as a manifestation of the known subtlety 
of holography for extremal correlators.

\setcounter{page}{0}
\setcounter{tocdepth}{2}

\newpage

\tableofcontents

\setcounter{footnote}{0}

\linespread{1.1}
\parskip 4pt

{}~
{}~

%%%%%%%%%%%%%%%%%%%%%%%%%%%%%%%%%%%%%%%
%%%%%%%%%%%%%%%%%%%%%%%%%%%%%%%%%%%%%%%
\section{Introduction}
%%%%%%%%%%%%%%%%%%%%%%%%%%%%%%%%%%%%%%%
%%%%%%%%%%%%%%%%%%%%%%%%%%%%%%%%%%%%%%%

The AdS/CFT correspondence \cite{Maldacena:1997re,Witten:1998qj,Gubser:1998bc}
and the discovery of integrability \cite{Minahan:2002ve,Beisert:2003tq,Beisert:2010jr} suggest that
it may be possible to solve the planar limit of four-dimensional ${\cal N}=4$ super 
Yang-Mills (SYM) theory. 
A complete solution to a conformal field theory is provided by the spectrum of
anomalous dimensions along with the three-point functions of primary operators.  
All higher order correlation functions can then, in principle, be obtained using the operator product expansion. 
The three-point functions are determined by the anomalous dimensions and the structure constants of the operator product 
expansion, both of which depend on the coupling constant of the theory. 

Using the AdS/CFT correspondence, correlation functions of
single-trace local gauge invariant operators can be computed at strong coupling.
This is achieved by inserting closed string vertex operators in the path integral for 
the string partition function. These vertex operators scale exponentially with the 
energy and the quantum conserved charges of the corresponding string states. 
When the conserved charges are of the order of the string tension, a saddle point approximation 
can be used to compute the string path integral.
This gives results in the semiclassical limit of large tension \cite{Gubser:2002tv}. 
Correlation functions involving heavy states (states with large conserved charges)
are then dominated by semiclassical string trajectories. 
Calculations of two-point functions in the semiclassical approach were performed 
in \cite{Tseytlin:2003ac,Buchbinder:2010gg,Janik:2010gc,Buchbinder:2010vw}. The works
\cite{Zarembo:2010rr},\cite{Costa:2010rz} discussed the computation of three--point functions
involving two heavy states and one light state. In this approach, one ignores the backreaction of
the light state on the heavy one, allowing the use of the same semiclassical trajectory. The
coupling to the light state is obtained through a fluctuation calculation. The application of
these semiclassical techniques to heavy-heavy-light correlators in $\mathcal{N}=4$ SYM has by now been extensively 
explored \cite{Roiban:2010fe,Hernandez:2010tg,Ryang:2010bn,Georgiou:2010an,
Park:2010vs,Bak:2011yy,Arnaudov:2011wq,Hernandez:2011up,Bai:2011su,Escobedo:2011xw,Alday:2011pf,
Ahn:2011zg,Lee:2011fe,Arnaudov:2011ek,Georgiou:2011qk,Bozhilov:2011qf,Alday:2011ga,Caetano:2011eb,Bozhilov:2011zp,
Bissi:2011ha,Hernandez:2012zj}.

One work of particular interest to us is \cite{Bissi:2011dc}. This article adopted the approach of 
\cite{Zarembo:2010rr} to study the correlation functions of giant gravitons \cite{McGreevy:2000cw,Hashimoto:2000zp}. 
These are solutions
of the D3-brane action, considered as a probe on the $\AdS_5\times \Srm^5$ geometry, which wrap an $\Srm^3$ 
within either the $\AdS$ or $\Srm^5$ and are stabilised by angular momentum along one of the $\Srm^5$ directions.
These bulk states have been shown to have a dual gauge theory description in terms of Schur polynomials \cite{Corley:2001zk}. 

Given that giant gravitons are heavy, semiclassical objects, it  appears very natural to apply the methods
of \cite{Zarembo:2010rr} to the calculation of their correlation functions. 
In particular, the correlation function considered in \cite{Bissi:2011dc} is that of two giant gravitons and one 
light graviton (dual to a chiral primary state of dimension
$\Delta =J\ll\sqrt{N}$). This correlator is protected from quantum corrections. Surprisingly,
\cite{Bissi:2011dc} find a discrepancy between gauge theory computations and the bulk calculations performed
using the probe brane action. 
Given that in the appropriate limit the probe brane result reduces very naturally to the three point-like graviton 
correlation function, while the gauge theory result does not reduce to that of three chiral primaries, 
\cite{Bissi:2011dc} argued that the disagreement might hinge on the inability of the Schur polynomials to 
interpolate between giant and point-like gravitons. 

 In order to obtain a better understanding of this mismatch, in this article we extend the work of \cite{Bissi:2011dc}
in two ways. We go beyond the probe analysis by calculating 3-point correlation functions using the fully backreacted
bubbling (LLM) geometry corresponding to the Schur polynomials in question. Furthermore, we compute \emph{non-extremal}
3-point functions using the semiclassical Born-Infeld analysis as well as the bubbling geometry. In the non-extremal
case, all three approaches are in perfect agreement, indicating that the Schur polynomials do provide a valid description 
of the giant gravitons and that the discrepancy in \cite{Bissi:2011dc} is due to known subtleties when performing bulk 
calculations for extremal correlators.  

The computation of \cite{Bissi:2011dc} was carried out entirely in the half-BPS sector. 
In the field theory, operators dual to giant gravitons of momentum $k$ wrapping the S$^3\subset$S$^5$ are
given by Schur polynomials in representation\footnote{Recall that Schur polynomials are labeled by
Young diagrams. Our notation is to list the row lengths. With this notation a Young diagram with a single 
row containing $k$ boxes is denoted $k$ while a Young diagram with a single column containing $k$ boxes is denoted $1^k$.} $1^k$ while operators dual to giant gravitons wrapping the S$^3\subset$AdS$_5$ are given by Schur polynomials in representation $k$. 
The light graviton is dual to ${\rm Tr}(Z^J)$. Notice that these are ``extremal'' 3 point functions, 
that is correlators in which the conformal dimension of one of the operators 
is precisely equal to the sum of the dimensions of the other two operators. 
It is well known that the computation of these correlators, using holography, is subtle \cite{D'Hoker:1999ea}. 
Supergravity fields will couple to linear combinations
of single and multiple trace BPS operators with the correct quantum numbers. 
It is simple to verify, using the free field approximation, that although operator mixing occurs, it is 
generically suppressed by powers of $1/N$. Extremal correlators are the exception. 
The contribution to extremal 3 point functions of multitrace operators is
enhanced and of the same order in $N$ as single traces. 
The origin of this enhancement can be traced back to the fact that the correlator effectively factorizes into a product of two point functions\footnote{We focus on three-point functions in this article. If one considers $n$ point functions with $n>3$, the same factorization can occur for next-to-extremal correlators and more generally, for near extremal correlators. The computation of this whole class of correlators using holography is subtle \cite{D'Hoker:2000dm,AdSreview}
and we expect that the comments we make in this article are valid for this whole class of correlators.}. In the dual gravitational description for these extremal correlators, the bulk supergravity coupling constants vanish but there is a boundary interaction that gives a nonvanishing result. This non-vanishing result agrees with field theory computations. 
Clearly, the gravitational computations of extremal correlators are rather subtle.

Further extensive comparison of extremal versus non-extremal correlators and their
holography has been carried out in \cite{Skenderis:2007yb},
in the half BPS sector of the theory, corresponding to the LLM geometries \cite{Lin:2004nb}.
These geometries are dual to Schur polynomials and superpositions of Schur polynomials.
They thus provide a natural setting within which we can reconsider the computation of \cite{Bissi:2011dc}.
Consider the vacuum expectation value (vev) of an operator with dimension $\Delta$
and ${\cal R}$-charge $J$. The case $J<\Delta$ corresponds to non-extremal correlators.
These correlators are not sensitive probes of the dual geometry and at large $N$ the results
depend only weakly on specifics of the distribution of Schur polynomials (see App. \ref{asympt} for a discussion 
of this fact using asymptotic representation theory). 
In sharp contrast to this, the vevs of maximally charged scalars depend on the
detailed Schur polynomial distribution.
This is precisely because the corresponding correlators are extremal \cite{Skenderis:2007yb}.
Thus, even at the leading order in $N$ these vevs are sensitive to the precise details
of the Schur polynomial distribution. These details are not captured by the (regular) supergravity
solution \cite{Skenderis:2007yb}, so its not surprising that a semiclassical computation is not
able to reproduce these correlators.

The gravitational computation in \cite{Bissi:2011dc} was carried out by
varying the Euclidean D-brane actions with respect to the supergravity
fluctuations corresponding to the light graviton. 
These fluctuations were then evaluated on appropriately Wick-rotated giant graviton solutions.
In this article we will compute these vevs corresponding to extremal correlators using the
holography formalism of \cite{Skenderis:2007yb}. 
The supergravity solutions we need are the LLM solutions \cite{Lin:2004nb}
which correspond to the exact geometry (including backreaction) dual to a giant graviton.
Our gravitational computation gives an exact agreement with the field theory computation
carried out using Schur polynomials.

To further test the idea that the mismatch between the semiclassical probe computation
and the field theory was due to subtleties related to extremal correlators, we consider
the computation of non-extremal correlators. 
In this case, the field theory, LLM and probe computations are in complete agreement.

The plan of the paper is as follows: In the next section we will first review the field theory
results of \cite{Bissi:2011dc} and then compute the same quantities using holography 
of the LLM geometries, finding complete agreement with the field theory results. We will also comment on
the failure of the semiclassical probe approach to reproduce these results. In section 
3 we will compute non-extremal correlators in three different ways: using field theory, LLM 
holography and a semiclassical probe. For all these cases we will find complete agreement.
The appendix contains a discussion of the insensitivity of non-extremal correlators on the
precise Young diagrams labelling the Schur polynomials.

%%%%%%%%%%%%%%%%%%%%%%%%%%%%%%%%%%%%%%%
%%%%%%%%%%%%%%%%%%%%%%%%%%%%%%%%%%%%%%%
\section{Extremal Correlators}
%%%%%%%%%%%%%%%%%%%%%%%%%%%%%%%%%%%%%%%
%%%%%%%%%%%%%%%%%%%%%%%%%%%%%%%%%%%%%%%

 In this section we focus on extremal correlators of giant gravitons with chiral primary operators, which we 
compute first using gauge theory techniques and then using LLM methods. Unlike the probe brane calculation
of \cite{Bissi:2011dc}, the LLM computation will be shown to yield complete agreement with the gauge theory. 
 
%%%%%%%%%%%%%%%%%%%%%%%%%%%%%%
\subsection{Gauge theory computation}
%%%%%%%%%%%%%%%%%%%%%%%%%%%%%%

Since the field theory computation has been carried out in \cite{Bissi:2011dc},
we will here simply summarize our notation and recall the result.
For references on the computations of correlation functions of Schur polynomials, 
we strongly recommend \cite{Corley:2001zk,Corley:2002mj}.

Recall that a giant graviton wrapping an S$^3\subset$ AdS$_5$ is dual to a Schur polynomial labeled by a
Young diagram with a single row containing order $N$ boxes, that is $\chi_{(k)}(Z)$, $k=O(N)$.
A giant graviton wrapping an S$^3\subset$ S$^5$ is dual to a Schur polynomial labeled by a
Young diagram with a single column containing order $N$ boxes, that is $\chi_{(1^k)}(Z)$, $k=O(N)$.
 
We will compute all correlators in ${\cal N}=4$ SYM theory on R$\times$S$^3$. All matrices
below are the s-wave of the KK expansion of the matrix field. Further, all correlators we compute
are equal-time correlators. A useful reference with further background on the relation between the
field theory on R$^4$ and on R$\times$S$^3$ is \cite{Berenstein:2004kk}.

The three-point function corresponding to two giant gravitons wrapping an S$^3\subset$ AdS$_5$ and a light graviton of angular momentum $J$ is
\bea
\langle O^{J,J}\rangle_{(k)} = { \langle \chi_{(k)}(Z)^\dagger\, {\rm Tr} (Z^J)\chi_{(k-J)}(Z)\rangle \over
   \sqrt{\langle \chi_{(k)}(Z)^\dagger\, \chi_{(k)}(Z)\rangle
   \langle {\rm Tr} (Z^J)^\dagger \, {\rm Tr} (Z^J)\rangle
   \langle \chi_{(k-J)}(Z)^\dagger \chi_{(k-J)}(Z)\rangle}}\;.
\label{FthreepointAdS}
\eea
We take $J$ to be of order 1.
When computing this three-point function using holography, it will be useful to write it 
as a one-point function of ${\rm Tr}(Z^J)$ in a particular state. 
The state we use is a superposition of two giants
$$
  |G\rangle =\left({\chi_{(k)}(Z)\over \sqrt{\langle \chi_{(k)}(Z)^\dagger\, \chi_{(k)}(Z)\rangle}}+{\chi_{(k-J)}(Z)\over \sqrt{\langle \chi_{(k-J)}(Z)^\dagger \chi_{(k-J)}(Z) \rangle}}
                \right)|0\rangle\, .
$$
Then, the correlator (\ref{FthreepointAdS}) is computed by the one point function
\be \nonumber
 \langle G|{{\rm Tr}(Z^J)\over\sqrt{\langle {\rm Tr} (Z^J)^\dagger \, {\rm Tr} (Z^J)\rangle}}|G\rangle \;.
\ee
In the limit we are considering ($N\rightarrow \infty$ with $k/N$ fixed and $J$ small), the 
value of this correlator is \cite{Bissi:2011dc}:
\bea \label{FthreepointAdSAnswr}
\langle O^{J,J}\rangle_{(k)} = {1\over\sqrt{J}}\left( 1+{k\over N}\right)^{\frac{J}{2}}\;.
\eea
We are also interested in the three-point function corresponding to two giant gravitons 
wrapping an S$^3\subset$ S$^5$ and a light graviton of angular momentum $J$, which 
is \cite{Bissi:2011dc}\footnote{We have dropped an irrelevant phase in writing this result.}
\bea
\langle O^{J,J}\rangle_{(1^k)} &&= { \langle \chi_{(1^k)}(Z)^\dagger\, {\rm Tr} (Z^J)\chi_{(1^{k-J})}(Z)\rangle \over
   \sqrt{\langle \chi_{(1^k)}(Z)^\dagger\, \chi_{(1^k)}(Z)\rangle
   \langle {\rm Tr} (Z^J)^\dagger \, {\rm Tr} (Z^J)\rangle
   \langle \chi_{(1^{k-J})}(Z)^\dagger \chi_{(1^{k-J})}(Z)\rangle}}\cr
   &&= {1\over\sqrt{J}}\left( 1-{k\over N}\right)^{\frac{J}{2}}\;.
\label{FthreepointS}
\eea
There is no $t$ dependence in these correlators because we have chosen to compute equal time correlators.
In this case we can employ a zero dimensional complex matrix model for the computation.

%%%%%%%%%%%%%%%%%%%%%%%%%%%%%%%%%%%%%%%
\subsection{LLM Computation}
%%%%%%%%%%%%%%%%%%%%%%%%%%%%%%%%%%%%%%%

According to the AdS/CFT correspondence, correlation functions can be computed in the strong coupling limit of 
${\cal N}=4$ SYM theory using the dual type IIB supergravity. 
In \cite{Skenderis:2007yb}, a powerful general approach to holography in the LLM backgrounds has been developed,
which allows the computation of correlation functions involving the gauge theory operators that create the LLM 
geometries. This is thus the correct framework to holographically compute (\ref{FthreepointAdSAnswr}) and (\ref{FthreepointS}). 
Note that, apart from correlators involving only local operators (as in our case), this approach has been successfully used to compute 
correlators involving surface operators \cite{Drukker:2008wr} and Wilson loops in large representations \cite{Gomis:2008qa}. 

To perform our computation, we need to construct the geometry dual to a superposition of two giant gravitons.
Since this is a half-BPS state in ${\cal N}=4$ SYM theory, it has a description in terms
of one-dimensional free fermions in an external potential. We will describe the two-dimensional phase space
density corresponding to this free fermion state. This is then to be identified with the defining density function
of the bubbling supergravity solution (see \cite{Lin:2004nb,Skenderis:2007yb} for further details). Having constructed
the LLM geometry, our desired correlation functions are then given by one-point functions of chiral primary operators in 
this background. What follows is a very quick review of the relevant material in \cite{Skenderis:2007yb}.

The second-quantized fermion field is
\bea
  \Psi (z,z^*,t)=\sum_{l=0}^\infty \hat{C}_l e^{-i(l+1)t}\Phi_l(z,z^*)\;,
\eea
with the fermionic oscillators obeying the standard anticommutators $\{ \hat{C}_l,\hat{C}_m^\dagger\} =\delta_{lm}$. 
The mode functions of the field are the orthonormal wave functions of the lowest Landau level
\bea
  \Phi_l(z,z^*)=\sqrt{2^{l+1}\over\pi\, l!}z^l e^{-zz^*}\;.
\eea
Since we have $N$ fermions the second quantized free fermion field obeys
\bea
  \int dzdz^* \Psi^\dagger (z,z^*,t)\Psi (z,z^*,t) =\sum_{l=0}^\infty \hat{C}_l^\dagger \hat{C}_l = N\;.
\eea
The Schur polynomials correspond to energy eigenstates of the $N$ fermion system.
The energies of the free fermions corresponding to Schur polynomial $\chi_R (Z)$ are
$$ E_i=r_i+N-i+1 \, \qquad i=1,...,N,$$
where $r_i$ is the number of boxes in the $i$th row of the Young diagram $R$. 
Using this interpretation it is straightforward to see that the Schur polynomial $\chi_R(Z)$ corresponds to the state
\bea
  |\chi_R\rangle\equiv \hat{C}_{N-1+r_1}^\dagger\hat{C}_{N-2+r_2}^\dagger \cdots \hat{C}_{1+r_{N-1}}^\dagger\hat{C}_{r_N}^\dagger |0\rangle\;.
\eea
Given this state, define
$$
  U_{lm}=\langle \chi_R|\hat{C}_l^\dagger \hat{C}_m|\chi_R\rangle
$$
from which we obtain the density function that determines the LLM geometry as
\bea
  \rho={1\over 2}\sum_{l,m} (z^*)^l z^m \sqrt{2^{2+l+m}\over \pi^2\, l!\, m!}e^{-2zz^*}U_{lm}\;.
  \label{LLMdensity}
\eea
As a consequence of the fact that the second quantized fermion field is normalized to $N$, 
these densities are all normalized so that
\bea
  \int_0^{2\pi} d\phi\int_0^\infty \, r dr\, \rho = {N\over 2}\;.
  \label{normed}
\eea
The computation of holographic one point functions now amounts to computing the integral\cite{Skenderis:2007yb}
\bea
  \langle G|{{\rm Tr} (Z^J)\over \sqrt{JN^J}}|G \rangle =  {N\over \sqrt{J}}\int r^{J}\rho\, e^{iJ\phi}\, r dr\, d\phi\;.
  \label{holone}
\eea
Our normalization conventions are chosen to match the normalizations used in the field theory
computation of the previous section. These differ slightly from the conventions of \cite{Skenderis:2007yb}. In particular,
we normalize the two point function of the light graviton to $1$ while \cite{Skenderis:2007yb} normalizes 
this to $N^2(J-1)(J-2)^2/\pi^4$. Thus, we rescale (3.60) of \cite{Skenderis:2007yb} by ${\pi^2}/N(J-2)\sqrt{J-1}$.
Also, our fields have been rescaled so that eigenvalues of $Z$ (to be identified with the free fermion coordinate $z$) 
run from $-1$ to $1$.
The supergravity coordinate $w$ is related to the free fermion coordinate $z$ of \cite{Skenderis:2007yb} by
\bea
   |z|=\sqrt{N\over 2}|w|\;.
\eea
Finally, as a consequence of the fact that we compute equal time correlators, we have removed the phase $e^{-iJt}$ which appears in (3.60) of \cite{Skenderis:2007yb}.

%%%%%%%%%%%%%%%%%%%%%%%%%%%%%
\subsubsection{AdS giant graviton}
%%%%%%%%%%%%%%%%%%%%%%%%%%%%%

In this case the Schur polynomials that participate are in the representations $(k)$ and $(k-J)$, and
\bea
    |G\rangle =|\chi_{(k)}\rangle +|\chi_{(k-J)}\rangle\;.
    \label{AdSG}
\eea
In the fermionic picture, to obtain an AdS giant of angular momentum $k$ we excite the fermion in the topmost
energy level of the Fermi sea by $k$ units of energy. 
Writing the state $|G\rangle$ in terms of fermionic oscillators we have
\bea
  |G\rangle = (\hat{C}^\dagger_{N+k-1}+\hat{C}^\dagger_{N+k-J-1})\hat{C}^{\dagger}_{N-2}\cdots \hat{C}^{\dagger}_{0}|0\rangle \;.
\eea
It is now simple to compute
\bea
  U_{lm}^G &&= \langle G|\hat{C}_l^\dagger \hat{C}_m|G \rangle\cr
        \cr
  &&=1 \qquad l=m=0,\, 1,\, 2,\cdots,N-2,\, \cr
  &&=1 \qquad l,m=N+k-1,\, N+k-J-1,\cr
  &&=0\qquad {\rm otherwise}
\eea
We can use this to compute $\rho$ and then the one point function using (\ref{holone}).
As a consequence of the integral over $\phi$ in (\ref{holone}), we see that only the term with coefficient 
$e^{-iJ\phi}$ in $\rho$ contributes to the one-point function. 
Thus, the only term in the density we need is ($z=re^{i\phi}$)
\bea
  \rho = {1\over \pi}(2z^*z)^{N+k-1-{J\over 2}}e^{-iJ\phi}{1\over \, \sqrt{(N+k-1)!(N+k-J-1)!}}e^{-2zz^*}
\eea
We now have\footnote{after inserting the rescaled variable $|w|$.}
\bea
  \langle G|{{\rm Tr} (Z^J)\over\sqrt{JN^{J}}}|G \rangle &&= 
       {N\over\sqrt{J}}
       {2N^{-J/2}\over\sqrt{(N+k-1)!(N+k-J-1)!}}
       \int_0^\infty (Nr^2)^{N+k-1}e^{-Nr^2}\, rdr\cr
  &&={1\over \sqrt{J}} \left(1+{k\over N}\right)^{J\over 2}\;.
\eea
This is in perfect agreement with the result (\ref{FthreepointAdSAnswr}) obtained using Schur polynomials.

%%%%%%%%%%%%%%%%%%%%%%%%%%%%%
\subsubsection{Sphere giant graviton}
%%%%%%%%%%%%%%%%%%%%%%%%%%%%%

In this case, the Schur polynomials that participate are in the representation $(1^k)$ and $(1^{k-J})$, and
\bea
    |G\rangle =|\chi_{(1^k)}\rangle +|\chi_{(1^{k-J})}\rangle\;.
    \label{SG}
\eea
In the fermionic picture, to obtain a sphere giant of angular momentum $k$ we create holes  
in the Fermi sea, by exciting the fermions in the $k$ top most energy levels of the Fermi sea by one 
unit of energy each. 
Writing the state $|G\rangle$ in terms of fermionic oscillators we have
\bea
  |G\rangle =(\hat{C}^\dagger_{N-k+J-1}+\hat{C}^{\dagger}_{N-k-1})\prod_{i=0,\ne N-k-1,\ne N-k+J-1}^N\hat{C}^\dagger_{i}|0\rangle
\eea
It is now simple to repeat the steps of the last section to obtain the only term in the density that gives a non-vanishing contribution 
\bea
  \rho = {1\over \pi}(2z^*z)^{N-k-1+{J\over 2}}e^{-iJ\phi}{1\over \, \sqrt{(N-k-1)!(N-k+J-1)!}}e^{-2zz^*}
\eea
such that the final answer for the correlator is
\bea
  \langle G|{{\rm Tr} (Z^J)\over\sqrt{JN^{J}}}|G \rangle = {1\over\sqrt{J}} \left(1-{k\over N}\right)^{J\over 2}\;.
\eea
This is again in perfect agreement with the result (\ref{FthreepointS}) obtained using Schur polynomials.

%%%%%%%%%%%%%%%%%%%%%%%%%%%%%%%%%%%%%%%%%%%
\subsection{Semiclassical failure of the Born-Infeld Action}
%%%%%%%%%%%%%%%%%%%%%%%%%%%%%%%%%%%%%%%%%%%

We are now faced with a stark paradox: the field theory computation using Schur polynomials performed in \cite{Bissi:2011dc} and the dual LLM computation are in perfect agreement. They both disagree with the computation using the Born-Infeld action presented in \cite{Bissi:2011dc}.
One is tempted to interpret this as a failure of the methods of \cite{Zarembo:2010rr} when studying 
the correlation functions of giant gravitons.
This can't be correct though: there are computations where the semiclassical methods employing the
Born-Infeld action correctly reproduce the field theory correlators \cite{Giombi:2006de}
and even cases where the field theory, semiclassical Born-Infeld and holography using bubbling 
geometries all agree \cite{Drukker:2008wr}.
This seems to suggest that the discrepancy is not because the semiclassical method is breaking down, but rather, because the correlators studied are very subtle. Indeed, our goal in this section is to argue that the failure of the semi-classical Born-Infeld computation is a consequence of the fact that the correlators considered are extremal. Note that the correlators computed in \cite{Giombi:2006de,Drukker:2008wr} are not extremal.

Our intuition is driven by the holography for LLM geometries, worked out in detail in \cite{Skenderis:2007yb}. 
Recall that to obtain a non-singular LLM geometry the density function determining the solution must only take the values $\{0,1\}$.
The density function describing the supergravity solution can be obtained, as we have described above, by translating the combination 
of Schur polynomials into a fermion state, and then computing the density using (\ref{LLMdensity}). 
This density function does not take only the values $\{0,1\}$, so that the corresponding supergravity solution is singular \cite{Lin:2004nb}. 
It is however possible to write the densities dual to a linear combination of Schurs as a sum of theta functions plus some $1/N$ corrections.
We would expect that, in the semiclassical limit, these $1/N$ corrections can be discarded. 
By discarding these corrections we are able to map Schur polynomials to regular supergravity geometries.
However, as a consequence of the fact that we have dropped these $1/N$ corrections, we no longer have a 
unique Schur polynomial for each (regular) supergravity geometry. 
Thus, in the semiclassical limit in which we have a smooth geometry,
we can not resolve the specific Schur polynomial in the dual gauge theory.

As we discussed in the introduction, \cite{Skenderis:2007yb} has argued that vevs of neutral operators in a state 
corresponding to a Schur polynomial of dimension $n$ do not depend on the fine details of the choice of Schur polynomial 
(see also App. \ref{asympt} for a discussion in the general non-maximally charged case). 
These vevs correspond to non-extremal three-point functions. We have no reason to expect that a semiclassical 
computation of these correlators will not succeed.
In contrast to this, the  vevs of maximally charged operators necessarily involve extremal correlators. 
The $N$ scalings of these vevs depend sensitively on the specific Schur polynomial corresponding to the supergravity
density \cite{Skenderis:2007yb}. 
For these extremal correlators it is not surprising that a semiclassical analysis fails as it did in \cite{Bissi:2011dc}.

Our argument makes a concrete prediction that can be tested: the computation of non-extremal correlators using
the gauge theory, a semiclassical Born-Infeld analysis or a holographic analysis using bubbling 
geometries should all agree. In the next section we show that this is indeed the case.

%%%%%%%%%%%%%%%%%%%%%%%%%%%%%%%%%%%%%%%
%%%%%%%%%%%%%%%%%%%%%%%%%%%%%%%%%%%%%%%
\section{Non-extremal Correlators}
%%%%%%%%%%%%%%%%%%%%%%%%%%%%%%%%%%%%%%%
%%%%%%%%%%%%%%%%%%%%%%%%%%%%%%%%%%%%%%%

The goal of this section is to consider the computation of a class of non-extremal correlators 
using three different approaches: using field theory, again using LLM holography and, finally, using a 
semiclassical probe approach. The three computations are in complete agreement which lends support to the 
idea that the mismatch found in \cite{Bissi:2011dc} is a consequence of the fact that the correlator studied is extremal.

The correlators we study are three-point functions of two heavy chiral primary operators which are maximally charged
with a third light operator (introduced in \cite{Skenderis:2007yb}) which is also a scalar chiral primary operator, 
but is not maximally charged. 
As a consequence of the fact that we include an operator which is not maximally charged, 
these correlators are non-extremal.
When performing the LLM and probe computations, it is again useful to arrange the computation as a one-point function
in the background created by the heavy maximally charged operators.
In general, this third light operator is built using all six of the adjoint scalars appearing in the theory. 
Thus, although we can compute one-point functions of this operator in the background of the heavy maximally charged operator
using the free fermion approach (which is all we do in this article), we would not be
able to compute two or higher point functions of these operators \cite{Skenderis:2007yb}. 

Thanks to known non-renormalization theorems protecting three-point functions of single trace chiral
primary operators of ${\cal N} = 4$ SYM (see \cite{nr1} and especially \cite{deBoer} for a recent
very readable proof) it is expected that three-point functions of 
protected multi-trace operators are not renormalized either \cite{nr2,nr3}. For this reason we expect 
matching between the field theory (i.e. weak coupling) and gravitational (i.e. strong coupling) computations we perform. 
Our results confirm this expectation.

%%%%%%%%%%%%%%%%%%%%%%%%%%%%%%%%%%%%%%%
\subsection{Gauge theory computation}
%%%%%%%%%%%%%%%%%%%%%%%%%%%%%%%%%%%%%%%

The operators that we consider are normalized as in \cite{Skenderis:2007yb}. This normalization is fixed by requiring that
our operator has vanishing one point function in the conformal vacuum, in the limit of large $S^3$ radius the 
vacuum expectation value of the operator must reduce 
to the Coulomb branch answer \cite{Skenderis:2006di} 
and its three-point functions with charged operators correctly reproduce the $\mathcal{N}=4$ SYM result.
The field theory computation is performed on $R\times S^3$, with all operators in the correlator at equal time. 
The fields are all the s-wave component of the KK-reduction of $Z,Z^\dagger$ on the $S^3$.
In this case, we can carry out the computation using a zero-dimensional complex matrix model.
The methods we use to compute the relevant correlators have been developed in \cite{Koch:2008cm}.

%%%%%%%%%%%%%%%%%%%%%%%%%%%%%%%%%%%%%%%
\subsubsection{Neutral operator}
%%%%%%%%%%%%%%%%%%%%%%%%%%%%%%%%%%%%%%%

The first correlator we would like to study is
\bea
\langle O^{2,0}\rangle \equiv \langle \chi_R (Z)\chi_R(Z)^\dagger {\sqrt{2}\over N\sqrt{3}}: {\rm Tr} (ZZ^\dagger):\rangle
\eea
where $:(\,\,):$ denotes normal ordering as usual. 
We perform this normal ordering by subtracting out self contractions.
All operators must be normal ordered since we know that the general structure
of the chiral primary operator is $C_{i_1\cdots i_p}{\rm Tr}(\phi^{i_1}\cdots \phi^{i_p})$ with $C_{i_1\cdots i_p}$ a symmetric
traceless tensor and $\phi^{i}$ the six hermitian adjoint scalars of the theory. 
The fact that $C_{i_1\cdots i_p}$ is traceless implies that all self-contractions vanish.

The correlator of interest is computed, up to normalization, by
\bea
{  \int \left[ dZdZ^\dagger\right] \chi_R (Z)\chi_R(Z)^\dagger {\rm Tr} (ZZ^\dagger) e^{-{\rm Tr} (ZZ^\dagger)} \over 
   \int \left[ dZdZ^\dagger\right] \chi_R (Z)\chi_R(Z)^\dagger e^{-{\rm Tr} (ZZ^\dagger)}}\;.
\eea
Consider the Schwinger-Dyson equation:
\bea
  0=\int \left[ dZdZ^\dagger\right] {d\over dZ_{ij}}\left( Z_{ij}\chi_R (Z)\chi_R(Z)^\dagger e^{-{\rm Tr} (ZZ^\dagger)}\right)\;.
\eea
We choose\footnote{i.e. $R$ is a Young diagram with $k$ boxes} $R\vdash k$ but 
otherwise leave $R$ unspecified. We will see that the result depends only on $k$, so that this
computation captures the answer for both the sphere and AdS giants.
Using the fact that $Z_{ij}{d\over dZ_{ij}}$ acting on an operator built from $Z$s just counts how many $Z$s there are in 
the operator, we find
\bea
  \langle \chi_R (Z)\chi_R(Z)^\dagger {\rm Tr} (ZZ^\dagger)\rangle &=& 
  \langle \left( Z_{ij}{d\over dZ_{ij}}\chi_R (Z)\right) \chi_R(Z)^\dagger \rangle
  +\langle \left({d\over dZ_{ij}}Z_{ij}\right)\chi_R (Z) \chi_R(Z)^\dagger \rangle\cr
  &=& (k+N^2)\langle \chi_R (Z) \chi_R(Z)^\dagger \rangle
\eea
so that
\bea
{ \langle \chi_R (Z)\chi_R(Z)^\dagger {\rm Tr} (ZZ^\dagger)\rangle\over \langle \chi_R (Z)\chi_R(Z)^\dagger \rangle}
=N^2+k\;.
\eea
To get the normal ordered answer, subtract the single contraction to find
\bea
  :{\rm Tr} (ZZ^\dagger): = {\rm Tr} (ZZ^\dagger) - N^2
  \label{normalorder}
\eea
so that our final result is
\bea
\langle O^{2,0}\rangle_{(k)} = \langle O^{2,0}\rangle_{(1^k)} =
{ \langle \chi_R (Z)\chi_R(Z)^\dagger {\sqrt{2}\over N\sqrt{3}}:{\rm Tr} (ZZ^\dagger):\rangle\over \langle \chi_R (Z)\chi_R(Z)^\dagger \rangle}
= \sqrt{\frac{2}{3}}{k\over N}\;.
\eea

%%%%%%%%%%%%%%%%%%%%%%%%%%%%%%%%%%%%%%%
\subsubsection{Charged operators}\label{chargedop}
%%%%%%%%%%%%%%%%%%%%%%%%%%%%%%%%%%%%%%%

The second type of correlators we study in the field theory are
$$
\langle O^{p,p-2}\rangle =  {\langle \sqrt{2\over p+1}{1\over N}:{\rm Tr}(Z^\dagger Z^{p-1}):\chi_R(Z)\chi_{R^+}(Z)^\dagger\rangle\over
   \sqrt{\langle\chi_{R^+}(Z)\chi_{R^+}(Z)^\dagger\rangle} \sqrt{\langle\chi_{R}(Z)\chi_{R}(Z)^\dagger\rangle} }
$$
where $R\vdash k$ and $R^+\vdash k+n-1$ and we will study the cases $p=3,4$.
The values of $\langle O^{p,2-p}\rangle$ are also easily obtained by complex conjugation.
Consider the Schwinger-Dyson equation:
$$
  0=\int \left[ dZdZ^\dagger\right] {d\over dZ_{ij}}\left( (Z^n)_{ij}\chi_R (Z)\chi_{R^+}(Z)^\dagger e^{-{\rm Tr} (ZZ^\dagger)}\right)\;.
$$
It implies that (subtracting out self contractions)
$$
\langle \chi_R (Z)\chi_{R^+}(Z)^\dagger :{\rm Tr} (Z^n Z^\dagger):\rangle = 
\langle \left( (Z^n)_{ij}{d\over dZ_{ij}}\chi_R (Z)\right) \chi_{R^+}(Z)^\dagger \rangle\;.
$$
Thus, we need to evaluate
\bea
(Z^n)_{ij}{d\over dZ_{ij}}\chi_R (Z) &=&
                        {1\over (k-1)!}\sum_{\sigma\in S_k}\chi_R (\sigma)
                        Z^{i_1}_{i_{\sigma(1)}}\cdots Z^{i_{k-1}}_{i_{\sigma(k-1)}}(Z^n)^{i_k}_{i_{\sigma(k)}}\cr
&=&{1\over (k-1)!}\sum_{\sigma\in S_k}\sum_{T\vdash\, k+n-1}\chi_R(\sigma)\chi_T(\sigma (k,k+1,\cdots,k+n-1)\,)\chi_{T}(Z)\cr
&=& {k\over d_R}\sum_{T\vdash\, k+n-1} {\rm Tr}(P_{T\to R} (k,k+1,\cdots,k+n-1)\,)\chi_{T}(Z)\;,
\eea
where $P_{T\to R}$ is a projection operator which projects us to $R$ after restricting to the $S_k$ subgroup and
is zero if $T$ does not subduce $R$ after restricting to the $S_k$ subgroup. 
The trace is over the carrier space of $T$. 
To obtain the final line above we have used the fundamental orthogonality relation to sum over $\sigma$.
Thus, we have
$$
\langle \chi_R (Z)\chi_{R^+}(Z)^\dagger :{\rm Tr} (Z^n Z^\dagger):\rangle = 
{k\over d_R} {\rm Tr}(P_{R^+\to R} (k,k+1,\cdots,k+n-1)\,)\langle \chi_{R^+} (Z) \chi_{R^+}(Z)^\dagger \rangle\;.
$$

For the symmetric representation $R=(k)$ and $R^+= (k+n-1)$. These irreducible representations are one-dimensional, so the
projector is trivial. Further, all group elements are represented by $1$. Thus we find  
$$
  {\langle :{\rm Tr}(Z^\dagger Z^n):\chi_{(k)}(Z)\chi_{(k+n-1)}(Z)^\dagger\rangle\over
   \sqrt{\langle\chi_{(k+n-1)}(Z)\chi_{(k+n-1)}(Z)^\dagger\rangle} \sqrt{\langle\chi_{(k)}(Z)\chi_{(k)}(Z)^\dagger\rangle} }
  =k\left(1+{k\over N}\right)^{n-1\over 2}\;.
$$

For the antisymmetric representation we have $R=(1^k)$ and $R^+= (1^{k+n-1})$. Again, these irreducible representations are 
one dimensional so again, the projector is trivial. In this representation group elements are $\Gamma (\sigma )=(-1)^{\epsilon(\sigma)}$
with $\epsilon(\sigma)$ the parity of the permutation. Thus we find
$$
  {\langle :{\rm Tr}(Z^\dagger Z^n):\chi_{(1^k)}(Z)\chi_{(1^{k+n-1})}(Z)^\dagger\rangle\over
   \sqrt{\langle\chi_{(1^{k+n-1})}(Z)\chi_{(1^{k+n-1})}(Z)^\dagger\rangle} \sqrt{\langle\chi_{(1^k)}(Z)\chi_{(1^k)}(Z)^\dagger\rangle} }
  =(-1)^{n-1\over 2} \, k\, \left(1-{k\over N}\right)^{n-1\over 2}\;.
$$
The phase $(-1)^{n-1\over 2}$ depends on our conventions and can easily be removed. For this reason we will drop this phase below.

\noindent
Reinstating the appropriate normalizations, we find
\bea
\langle O^{3,1}\rangle_{(k)} = \sqrt{\frac{1}{2}}{k\over N}\sqrt{1+{k\over N}}\;,
\qquad
\langle O^{3,1}\rangle_{(1^k)} = \sqrt{\frac{1}{2}}{k\over N}\sqrt{1-{k\over N}} \;,
\eea
and
\bea
\langle O^{4,2}\rangle_{(k)} = \sqrt{\frac{2}{5}}{k\over N}\left( 1+{k\over N}\right)\;,
\qquad
\langle O^{4,2}\rangle_{(1^k)} = \sqrt{\frac{2}{5}}{k\over N}\left( 1-{k\over N} \right)\;.
\eea
The results for $\langle O^{3,-1}\rangle$ and $\langle O^{4,-2}\rangle$ follow from these by complex conjugation.

%%%%%%%%%%%%%%%%%%%%%%%%%%%%%%%%%%%%%%%
\subsection{LLM Computation}
%%%%%%%%%%%%%%%%%%%%%%%%%%%%%%%%%%%%%%%

The vev of the neutral scalar dual to $O^{2,0}$ is given by \cite{Skenderis:2007yb}
\bea
  \langle G| {\sqrt{2}\over N\sqrt{3}} :{\rm Tr} (ZZ^\dagger): |G \rangle
             = {\sqrt{2}N\over \sqrt{3}} \int \rho (r^2-{1\over 2})rdrd\phi\;.
\label{LLMO20}
\eea
By explicit computation, we have verified that the vevs are the same for LLM states 
$|G\rangle =|\chi_{(k)}\rangle$ or $|G\rangle =|\chi_{(k-J)}\rangle$
as well as for $|G\rangle =|\chi_{(1^k)}\rangle$ or $|G\rangle =|\chi_{(1^{k-J})}\rangle$, because
in the limit we study we cannot distinguish between $k$ and $k-J$.
In the above equation $\rho$ is given by the distribution corresponding to the Schur ($\rho_{\chi_{(k)}}$) 
minus the distribution corresponding to the vacuum ($\rho_{\Omega}$)
\begin{equation}
\rho=\rho_{\chi_{(k)}}-\rho_{\Omega}=\frac{1}{\pi}e^{-N r^2}\left(\frac{(Nr^2)^{k}}{k!}-1\right)\;.
\end{equation}
Notice that this is completely equivalent to the subtraction (\ref{normalorder}) performed in the field theory.
Inserting this density into \eqref{LLMO20} gives perfect agreement with the gauge theory computation
\bea
  \langle O^{2,0}\rangle =  \langle G| {\sqrt{2}\over N\sqrt{3}} :{\rm Tr} (ZZ^\dagger): |G \rangle={\sqrt{2}\over\sqrt{3}}{k\over N}.
\eea
This result depends only on the number of boxes $k$ in the Schur polynomial and for this reason the result
for the sphere giant and the AdS giant are identical.

Next, consider the vev of $O^{3,1}$. This is computed 
using\footnote{This differs from formula (3.61) of \cite{Skenderis:2007yb}. Replacing $( r^2- 1 )$ by $r^2$ we would recover
(3.61) of \cite{Skenderis:2007yb}, up to normalization. 
The $-1$ implements the normal ordering. This is not needed in \cite{Skenderis:2007yb} since
they limit themselves to densities for which $\int d^2 w\, w\rho =0$. It is a simple matter to check that the density we consider
here does not satisfy this constraint.}
\bea
  \langle O^{3,1}\rangle ={N\over \sqrt{2}} \int \rho e^{i\phi}\,( r^2- 1 )\, rdrd\phi
  \label{modifiedST31}
\eea
Here $\rho$ is computed using the state $|G\rangle$ in (\ref{AdSG}) with $J=1$ for the AdS giant computation and by the state $|G\rangle$ in
(\ref{SG}) with $J=1$ for the sphere giant computation. It is now straight forward to find, for the AdS giant computation,
\bea
  \langle O^{3,1}\rangle ={2\sqrt{N}\over \sqrt{2}} \int_0^\infty \,
                   {e^{-Nr^2} (Nr^2)^{2k-2+2N\over 2}\over \sqrt{(k+N-2)!(k+N-1)!}}\,( r^2- 1 )\, rdr
                   =\sqrt{\frac{1}{2}}{k\over N}\sqrt{1+{k\over N}}
\eea
in complete agreement with the field theory result. Again, $\langle O^{3,-1}\rangle$ can be obtained by complex conjugation
of this result. The sphere giant result is also in complete agreement with the field theory.

The vev of $O^{4,2}$ is computed using\footnote{It is easy to check that the $-r^2$ in the next formula 
is the same subtraction that implements the normal ordering in the field theory. This is, up to normalization, 
(3.62) of \cite{Skenderis:2007yb}.}
\bea
  \langle O^{4,2}\rangle ={2N\over \sqrt{10}} \int \rho e^{i\phi}\,( r^4- r^2 )\, rdrd\phi\;.
  \label{modifiedST42}
\eea
Here $\rho$ is computed using the state $|G\rangle$ in (\ref{AdSG}) with $J=2$ for the AdS giant computation and by the state $|G\rangle$ in
(\ref{SG}) with $J=2$ for the sphere giant computation. It is now straightforward to find, for the AdS giant computation,
\bea
  \langle O^{4,2}\rangle ={2\sqrt{2}\over \sqrt{5}} \int_0^\infty \,
                   {e^{-Nr^2} (Nr^2)^{2k-2+2N\over 2}\over \sqrt{(k+N-3)!(k+N-1)!}}\,( r^2 - 1 )\, rdr
                   =\sqrt{\frac{2}{5}}{k\over N}\left( 1+{k\over N}\right)
\eea
in complete agreement with the field theory result. Again, $\langle O^{4,-2}\rangle$ can be obtained by complex conjugation
of this result. One also finds complete agreement for the sphere giant result.

%%%%%%%%%%%%%%%%%%%%%%%%%%%%%%%%%%%%%%%
\subsection{Probe Computation}
%%%%%%%%%%%%%%%%%%%%%%%%%%%%%%%%%%%%%%%

 In this section we turn to the computation of non-extremal correlators using the probe approximation.  
We closely follow \cite{Bissi:2011dc} and use the same notation for easy comparison. In particular,
we write the metric of global $\AdS_5\times \Srm^5$ as:
\be
\diff s^2=-\cosh^2\rho~\diff t^2+\diff \rho^2+\sinh^2\rho\,\diff\tilde{\Omega}_3^2+\diff\theta^2+\sin^2\theta\,\diff\phi^2
+\cos^2\theta\,\diff\Omega_3^2
\ee
where $\diff\tilde\Omega_3^2$ (parametrized by angles $\tilde\chi_i$) and $\diff\Omega_3^2$ (parametrized by
angles $\chi_i$) are volume elements of three-spheres inside $\AdS_5$ and $\Srm^5$ respectively. 

As discussed in \cite{Bissi:2011dc}, where the semiclassical methods of \cite{Zarembo:2010rr} for semiclassical strings
were generalized to D-branes, to compute the correlator of a giant graviton with a chiral primary operator one
needs to vary the  Euclidean D-brane action%
\bea
  S_{D3}=S_{DBI}+S_{WZ}={N\over 2\pi^2}\int d^4\sigma (\sqrt{\mathrm{det}(g_{MN}\partial_a X^M\partial_b X^N)}-iP\big[ C_4\big])
\eea
with respect to the supergravity fluctuation dual to the chiral primary operator
%$:{\rm Tr}(ZZ^\dagger):$ 
and evaluate the fluctuation on the Wick-rotated giant graviton solution, defined on the Poincar\'e patch of $\AdS_5$.

 In \cite{Bissi:2011dc}, this computation was performed only for extremal correlators, 
where the chiral primary is taken to be $O^{(J,J)}\sim\Tr Z^J$. The bulk scalar dual to this 
operator can be written as $s=s_\Delta Y^{(\Delta,\Delta)}$, where $s_\Delta$ is its bulk-to-boundary propagator and 
\be \label{YDD}
Y^{(\Delta,\Delta)}=\frac{\sin^\Delta\theta e^{i\Delta \phi}}{2^{\Delta/2}}
\ee
is the corresponding spherical harmonic on $\Srm^5$. In the present case we are 
interested in non-extremal correlators, involving the operators 
\be
O^{(p,p-2)}\sim~ :\Tr Z^{p-1}Z^\dagger: \;.
\ee
The corresponding harmonics on $\Srm^5$ can be found e.g. in \cite{Skenderis:2007yb} and for the first few low-dimension operators
are:
\be
\begin{split} \label{Ynonext}
Y^{(2,0)}&=\frac{1}{2\sqrt{3}}(3\sin^2\theta-1)\;,\\
Y^{(3,1)}&=\frac{\sqrt{3}}4 \sin\theta ~(2\sin^2\theta-1) ~e^{i\phi}\;,\\
Y^{(4,2)}&=\frac{1}{2\sqrt{10}}\sin^2\theta~(5\sin^2\theta-3)~e^{2 i \phi}\;.
\end{split}
\ee
As will become clear, the only differences between the calculations of \cite{Bissi:2011dc} and ours
will follow from making the above choices of spherical harmonics instead of (\ref{YDD}).

%%%%%%%%%%%%%%%%%%%%%%%%%%%%%
\subsubsection{Sphere giant graviton}
%%%%%%%%%%%%%%%%%%%%%%%%%%%%%

For this giant graviton, whose worldvolume is along $t\in \AdS_5$ and an $\Srm^3\in \Srm^5$, the brane ansatz 
is given by \cite{Grisaru:2000zn}
\be
\rho=0\;,\quad\sigma^0=t\;,\quad \phi=\phi(t)\;,\quad \sigma^i=\chi_i\;.
\ee
Substituting this ansatz into the D3-brane action one finds that the energy is minimized by
\be \label{minEnergy}
\cos^2\theta=\frac{k}{N}\; \quad \text{and}\quad \dot{\phi}=1\;,
\ee 
where the energy is $E_{min}=k$. In other words, the brane is moving at constant velocity along the $\phi$ angle,
with its energy depending on its latitude $\theta$ and reaching its maximum value at $\theta=0$. 

The above solution describes a time-dependent solution of the D3-brane action. As explained in 
\cite{Tsuji:2006zn,Zarembo:2010rr}, mapping to the Poincar\'e patch and performing appropriate 
Wick rotations transforms this solution to one starting and ending on the $\AdS$ boundary, which can
thus be identified with a two-point correlation function in $\mathcal{N}=4$ SYM theory via the standard
AdS/CFT dictionary. The three-point function with a light chiral primary can then be computed by considering
fluctuations of the supergravity fields that couple the solution to the bulk field dual to the chiral primary. 

After performing these transformations, the scalar bulk-to-boundary propagator from an arbitrary boundary 
point $x_B$ (taken to be very far from the giant graviton endpoints) to the location of the giant 
graviton at $\rho=0$ is simply   
\be
s_\Delta=\frac{\Delta+1}{N\Delta^\half 2^{2-\frac{\Delta}{2}}}\frac{\Rcal^\Delta}{\cosh^\Delta t} 
\ee
with $\Rcal=R/x_B^2$. Furthermore, after Wick rotation the condition $\dot\phi=1$ in (\ref{minEnergy}) is solved by 
$\phi=-i t$, which leads to a $t$-dependence of the harmonics (\ref{YDD},\ref{Ynonext}) when evaluated on the brane
worldvolume.

The details of the fluctuation calculation for this case can be found in \cite{Bissi:2011dc}. Generalizing slightly
to include the possibility of more general harmonics, we can write the variation of the DBI part of the action as:
\be
\delta S_{DBI}=\frac{N}{2}\cos^2\theta\int_{-\infty}^\infty \diff t~ Y^{(\Delta,J)}(t,\theta)\left[\frac{4}{\Delta+1}\partial_t^2
-\left(\frac{2\Delta(\Delta-1)}{\Delta+1}+8\Delta\sin^2\theta-6\Delta\right)\right] s_\Delta
\ee
and that of the Wess-Zumino part as:
\be
\delta S_{WZ}=-4N\sin\theta\cos^3\theta\int_{-\infty}^\infty \diff t ~s_\Delta \partial_\theta Y^{(\Delta,J)}(t,\theta)\;.
\ee
Performing the $t$ integrals for the relevant cases, we obtain:
\be
\begin{split}
\delta S(Y^{(2,0)})&=-\sqrt{\frac{2}{3}} ~(2\Rcal)^2 \cos^2\theta\quad \Rightarrow\quad 
\langle O^{(2,0)}\rangle_{(1^k)}=\sqrt{\frac23} \frac{k}{N}\;,\\
\delta S(Y^{(3,1)})&=-\half\sqrt{2}(2\Rcal)^3\cos^2\theta\sin\theta\quad\Rightarrow\quad
\langle O^{(3,1)}\rangle_{(1^k)}=\frac{1}{\sqrt2}\frac{k}{N}\sqrt{1-\frac{k}{N}} \;,\\
\delta S(Y^{(4,2)})&=-\sqrt{\frac{2}{5}} (2\Rcal)^4 \sin^2\theta\cos^2\theta \quad\Rightarrow\quad
\langle O^{(4,2)}\rangle_{(1^k)}=\sqrt{\frac25}\frac kN \left(1-\frac kN\right)\;.\\
\end{split}
\ee
We find that, in stark contrast to the extremal correlator calculations of \cite{Bissi:2011dc}, the 
semiclassical calculation of non-extremal giant graviton correlation functions is in perfect agreement
with the gauge theory and LLM computations.

%%%%%%%%%%%%%%%%%%%%%%%%%%%%
\subsubsection{AdS giant graviton}
%%%%%%%%%%%%%%%%%%%%%%%%%%%%

 Now we consider the dual giant graviton, wrapping $t$ as well as an $\Srm^3$ inside $\AdS_5$. The corresponding
ansatz is \cite{Grisaru:2000zn}:
\be
\rho=\text{const.}\;,\quad \sigma^0=t\;,\quad \sigma^i=\tilde{\chi}_i\;,\quad \phi=\phi(t)\;,\quad \theta=\frac{\pi}{2} \;.
\ee
{}From this ansatz, one finds that the minimum of the energy, $E_{min}=k$, is at
\be
\sinh^2\rho=\frac{k}{N}\quad \text{and} \quad \dot{\phi}=1\;. 
\ee
Thus the brane is still moving along the $\phi$ angle of the $\Srm^5$ with constant velocity but this time the
energy depends on the radial coordinate $\rho$ (and is unbounded, since the radius $\sinh\rho$ of the wrapped
sphere can become arbitrarily large). As before, mapping to the Poincar\'e patch and performing appropriate
Wick rotations leads to a solution that can be interpreted as a gauge theory two-point function. Coupling the
solution to a chiral primary by considering fluctuations of the supergravity fields will lead us to the three-point
functions of interest. 

 For this case, where the brane is not located at $\rho=0$, the required bulk-to-boundary propagator is more complicated:
\be
s_\Delta=\frac{\Delta+1}{4\sqrt{\Delta}N}\frac{\Rcal^\Delta}{(\cosh\rho\cosh t-\cos\vartheta\sin\phi_1\sinh\rho)^\Delta}\;,
\ee
where the $\Srm^3$ wrapped by the brane is spanned by the coordinates $\tilde\chi_i=\{\vartheta,\phi_1,\phi_2\}$. 
The fluctuation calculation for this case was also performed in \cite{Bissi:2011dc}. Independently, the DBI and Wess-Zumino 
parts are rather complicated, but summing them leads to the following simple result:
\be
\begin{split}
\delta S=&-\int_{-\infty}^{\infty}\diff t\int_0^{2\pi}\diff \phi_1\int_0^{2\pi}\diff\phi_2\int_0^{\pi/2}\diff\vartheta\\
&\times\frac{2^{\frac{\Delta}2}\sqrt{\Delta}\sqrt{\Delta+1}}{4\pi^2}\cos\vartheta\sin\vartheta\sinh^2\rho\frac{\Rcal^\Delta Y^{(\Delta,J)}(t)}{(\cosh\rho\cosh t-\cosh\vartheta\sin\phi_1\sinh\rho)^{\Delta+2}}\;.
\end{split}
\ee
Here we have generalized slightly to include the possibility of arbitrary $Y^{(\Delta,J)}(t)$, where as before
the $t$ dependence arises through $\phi=-it$ (and $\theta=\pi/2$ as required by the brane solution). 
To evaluate the integrals, we follow the steps of \cite{Bissi:2011dc} and perform a formal expansion in
the quantity $[\cos\vartheta\sin\phi_1\tanh\rho/\cosh t]$, after which the $\vartheta$ and $\phi_1$ integrations
are straightforward. Proceeding exactly as in \cite{Bissi:2011dc}, we are left  with the $t$ integral 
and resummation of the series:
\be
\delta S=-\frac{2^{\frac{\Delta}{2}-1}\sqrt{\Delta}}{\Gamma(\Delta+1)}\frac{\Rcal^\Delta}{\cosh^\Delta\rho}
\sum_{n=0}^{\infty}\frac{1}{2^{2n}}\frac{\Gamma(\Delta+2n+2)}{\Gamma(n+2)\Gamma(n+1)}\tanh^{2n+2}\rho
\int_{-\infty}^\infty \frac{Y^{(\Delta,J)}}{\cosh^{\Delta+2+2n}t}\;.
\ee
Evaluating this expression for the spherical harmonics of interest (\ref{Ynonext}), we obtain:
\be
\begin{split}
\delta S(Y^{(2,0)})&=-\sqrt{\frac23}(2\Rcal)^2\sinh^2\rho\quad \Rightarrow\quad  
\langle O^{(2,0)}\rangle_{(k)}=\sqrt{\frac23}\frac{k}{N}\;,\\
\delta S(Y^{(3,1)})& =-\frac{(2\Rcal)^3}{\sqrt{2}} \sinh^2\rho\cosh\rho \quad\Rightarrow\quad
\langle O^{(3,1)}\rangle_{(k)}=\frac{1}{\sqrt{2}}\frac kN\sqrt{1+\frac kN}\;,\\
\delta S(Y^{(4,2)})&=-\sqrt{\frac 25} (2\Rcal)^4\sinh^2\rho\cosh^2\rho \quad \Rightarrow \quad
\langle O^{(4,2)}\rangle_{(k)}=\sqrt{\frac 25}\frac kN\left(1+\frac kN\right)\;.
\end{split}
\ee
We again find perfect agreement with the gauge theory and LLM computations, which strongly
suggests that the semiclassical approach can safely be applied to any non-extremal correlators.

%%%%%%%%%%%%%%%%%%%%%%%%%%%%%%%%%%%%%%%
%%%%%%%%%%%%%%%%%%%%%%%%%%%%%%%%%%%%%%%
\section{Conclusions}
%%%%%%%%%%%%%%%%%%%%%%%%%%%%%%%%%%%%%%%
%%%%%%%%%%%%%%%%%%%%%%%%%%%%%%%%%%%%%%%

We have studied three-point functions for two giant gravitons and one light graviton. 
In the case that the computation involves extremal correlators in the field theory we have found perfect agreement
between the field theory and holography in LLM backgrounds. The results disagree with semiclassical 
computations using the Born-Infeld action. We have argued that this disagreement
is a consequence of the fact that one is computing extremal correlators. In support of this interpretation, we
have also considered three-point functions for two giant gravitons and one light graviton that involve
only non-extremal correlators in the field theory. In this case we find perfect agreement between the field theory,
LLM holography and semiclassical probe computations.

It is interesting to note that in \cite{Zarembo:2010rr} the semiclassical string calculation of an extremal three-point
function of chiral primaries was found to be in exact agreement with the corresponding field theory result. 
This is not unlike the computation \cite{nr1} of three-point functions of the chiral primary operators, 
for which correlators computed at strong coupling using supergravity were found to exactly agree at 
large $N$ with the free field approximation to the super Yang-Mills theory.
Upon closer examination of the computation of \cite{nr1}, one finds subtleties in the case 
of extremal three-point functions \cite{D'Hoker:1999ea}.
A relevant recent computation is reported in \cite{Klose:2011rm,Buchbinder:2011jr}.
These papers perform a semiclassical computation in light-cone gauge for the worldsheet theory, for the three-point functions of
BMN vertex operators, with all three operators heavy.
They find that for extremal correlators the saddle point localizes at the boundary $z=0$ in agreement with \cite{D'Hoker:1999ea}.
Further, in this case the action diverges and the correlator should be defined as a limit of the non-extremal one \cite{Buchbinder:2011jr}. We have found no such subtlety in the chiral primary computation of \cite{Zarembo:2010rr}, however
there is a known subtlety in obtaining the extremal chiral primary result as a limit of a computation where the heavy 
states are non-BPS. In \cite{Russo:2010bt} it was argued that the normalisation of the light operator should be 
adjusted appropriately in order for this limit to be smooth. It would clearly be interesting to study this issue further and 
determine whether a more refined semiclassical calculation could resolve the mismatch in the extremal three-point function. 

 Our comparison of the various methods in the non-extremal case was for certain specific classes of non-maximally-charged
operators. In particular, we focused for simplicity on operators of low dimension. It would be interesting to extend
these results to higher-dimension operators, as well as to operators of different types (e.g. involving the stress-energy
tensor of the theory). This would provide further support for our proposal that subtleties in the semiclassical approach
to correlation functions of giant gravitons appear only in the extremal case.

%%%%%%%%%%%%%%%%%%%%%%%%%%%%%%%%%%%%%%%
%%%%%%%%%%%%%%%%%%%%%%%%%%%%%%%%%%%%%%%
\section*{Acknowledgments}
%%%%%%%%%%%%%%%%%%%%%%%%%%%%%%%%%%%%%%%
%%%%%%%%%%%%%%%%%%%%%%%%%%%%%%%%%%%%%%%

We thank Shinji Hirano, Vishnu Jejjala, Charlotte Kristjansen, Sanjaye Ramgoolam and Donovan Young for useful discussions and comments. 
PC and RdMK are supported by the South African Research Chairs Initiative of the Department of Science and Technology and National Research Foundation.
KZ is supported by Danish Research Council grant ``Integrable Theories of Particles and Strings''.

\begin{appendix}

\section{Further comments on non-extremal correlators} \label{asympt}

A key idea of this article has been that non-extremal correlators depend only weakly on the Young diagram label $R$ of the Schur polynomial,
a fact that was explored in \cite{Skenderis:2007yb}.
In this Appendix, by making use of the asymptotic representation theory of the symmetric group\cite{thoma,vk}, we will make some precise
observations on this dependence. 

In section (\ref{chargedop}) we found the exact correlator
\bea
{\langle \chi_R (Z)\chi_{R^+}(Z)^\dagger :{\rm Tr} (Z^n Z^\dagger):\rangle\over
\langle \chi_{R^+} (Z) \chi_{R^+}(Z)^\dagger \rangle } = 
{k\over d_R} {\rm Tr}(P_{R^+\to R} (k,k+1,\cdots,k+n-1)\,) \;.
\label{insensitive}
\eea
The trace on the RHS in the above equation is taken over the carrier space of $R^+$.
The construction of the projection operators $P_{R^+\to R}$ has been considered in detail in \cite{dssi}.
These projectors can be written as polynomials in Jucys-Murphy elements $J_m$ with $m=k,k+1,...,k+n-1$
where in cycle notation
$$
  J_m=\sum_{i=1}^{m-1} (i,m) \;.
$$
$J_m$ takes values in the group algebra of $S_m$.
Thus, (\ref{insensitive}) is a sum over normalized characters\footnote{In the notation of \cite{vk}, a normalized character is
given by the usual character divided by the dimension of the representation. It is the normalized characters of the symmetric 
group $S_p$ that have a nice limit as $p\to\infty$\cite{vk}.} times powers of $k$. Notice that $k$ depends only on the number
of boxes in the Young diagram $R$ so that it is completely independent of the shape of $R$. Thus, the only dependence in 
(\ref{insensitive}) on the shape of the Young diagram is due to the normalized characters. In the remainder of this appendix
we summarize the relevant pieces of \cite{vk} which prove that the normalized characters depend only weakly on the
representation.

We will introduce the Frobenius coordinates $a_i$, $b_i$ for a Young diagram by means of an example. 
Look at the Young diagram:
$$
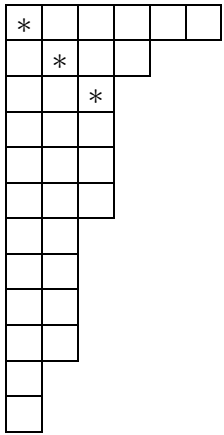
\young({*}{\,}{\,}{\,}{\,}{\,},{\,}{*}{\,}{\,},{\,}{\,}{*},{\,}{\,}{\,},{\,}{\,}{\,},{\,}{\,}{\,},{\,}{\,},{\,}{\,},{\,}{\,},{\,}{\,},{\,},{\,})
$$
Consider the boxes on the diagonal of the Young diagram, marked with a star in the figure above.
This is called the principal diagonal of the Young diagram.
There are a pair of coordinates $(a_i,b_i)$ for each starred box.
The $a_k$ are the number of boxes in row $k$ minus $k$ minus $1/2$. 
For the diagram shown $a_1=5.5$, $a_2=2.5$ and $a_3=0.5$. 
The definition of the $b_i$ are the same but we now use the columns, so that for the above diagram we have $b_1=11.5$, $b_2=8.5$ and $b_3=3.5$. 
Denote the total number of boxes in the Young diagram by $n$. Define
\bea
   \alpha_k ={a_k\over n},\qquad \beta_k ={b_k\over n} \;.
\eea
Thoma's theorem (see \cite{vk} for a very readable account) says that as $n\to\infty$ the value of a normalized character 
of permutation $\sigma$ in the representation with Frobenius coordinates $a_k$, $b_k$ is
$$
  \prod_{m\ge 2}(\sum_i \alpha_i^m + (-1)^{m+1}\sum_i \beta_i^m)^{\rho_m(\sigma )}
$$
where $\rho_m(\sigma)$ is the number of cycles of length $m$ in $\sigma$.

Now, in the limit $N\to\infty$ since $n$ is order $N$, shifting $l$ boxes around with $l/N\ll 1$, 
will not affect the value of the normalized character.
We have focused on representations with one long row or column.
One can shift $N^{1-\epsilon}$ boxes out of the long row (say) with $1\gg\epsilon >0$ and put them back into any number
of additional rows or columns, building up any shape, without changing the answer for the normalized character. 
This demonstrates that (\ref{insensitive}) is remarkably insensitive to the shape of the Young diagram.

\end{appendix}

%%%%%%%%%%%%%%%%%%%%%%%%%%%%%%%%%%%%%%%
%%%%%%%%%%%%%%%%%%%%%%%%%%%%%%%%%%%%%%%

\end{document}